\documentclass[12pt]{article}

\usepackage[latin1]{inputenc}
\usepackage{graphicx,color}

\makeatletter
\def\btt#1{\texttt{\@backslashchar#1}}%
\DeclareRobustCommand\bblash{\btt{\@backslashchar}}%
\makeatother

\begin{document}

\begin{center}
{\bf \large  LOW FRICTION  FLOWS OF LIQUIDS AT NANOPATTERNED
INTERFACES}

C\'ecile Cottin-Bizonne,
 Jean-Louis Barrat\footnote[1]{Author for correspondance:
 barrat@dpm.univ-lyon1.fr; Tel: 33 4 72 44 85 65 Fax: 33 4 72 43 26 48},Lyd\'eric Bocquet,
  Elisabeth Charlaix

\end{center}

\medskip

 \textsl{D\'epartement de Physique des Mat\'eriaux, CNRS and  Universit\'e
de Lyon, Bâtiment Léon Brillouin,  43 Boulevard du 11 Novembre,
69622 Villeurbanne Cedex}

\bigskip

\hrule

\bigskip
{\bf With the recent important development of microfluidic
systems, miniaturization of flow devices has become a real
challenge. Microchannels, however, are characterized by a large
surface to volume ratio, so that surface properties strongly
affect flow resistance in submicrometric devices. We present here
results showing that the concerted effect of wetting properties
and surface roughness may considerably reduce friction of the
fluid past the boundaries. The slippage of the fluid at the
channel boundaries is shown to be drastically increased by using
surfaces that are patterned at the nanometer scale. This effect
occurs in the regime where the surface pattern is partially
dewetted, in the spirit of the 'superhydrophobic' effects that
have been recently discovered at the macroscopic
scales\cite{quere2002}. Our results show for the first time that,
in contrast to the common belief, surface friction may be reduced
by surface roughness. They also open the possibility of a
controlled realization of the 'nanobubbles' \cite{tyrell2001} that
have long been suspected to play a role in interfacial slippage
\cite{vino95,degennes2002}.}

\begin{figure}[h]
\centering
\includegraphics[width=8cm]{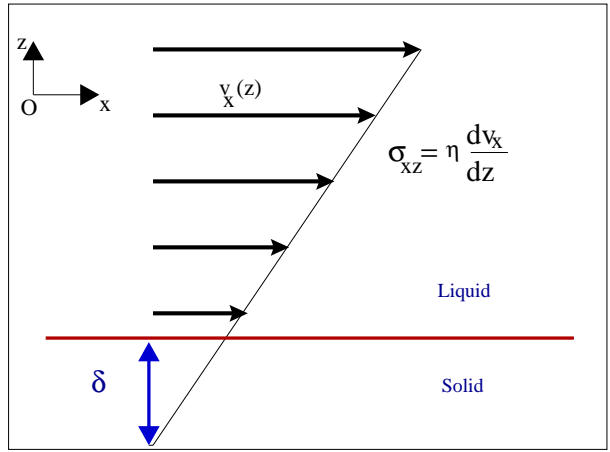}
\caption{Schematic definition of the slip length. The linear
velocity profile in the flowing, newtonian fluid (characterized by
a Newtonian viscosity $\eta$ and a constant shear stress
$\sigma_{xz}$) does not vanish at the solid boundary.
Extrapolation into the solid at a depth $\delta$ is necessary to
obtain a vanishing velocity, as assumed by the macroscopic
'no-slip' boundary condition.} \label{sliplengthdef}
\end{figure}

\newpage
The nature of the boundary condition for fluid flows past solid
surfaces is a subject of ancient interest
\cite{Schnell,Churaev84}, which has been revived recently by  a
large number of experiments and new theoretical approaches. The
possibility of investigating flows at small scales in a
quantitative manner, as opened by the development of nanoscale
measurements (Surface force apparatus or Atomic Force Microscope)
has allowed a number of experimental determinations of the 'slip
length' $\delta$  (see figure\ref{sliplengthdef}) that is used to
characterize this  boundary condition
\cite{Horn85,Georges93,Baudry2001,Craig2001,Granick2002,soumis}.
Optical techniques, such as fluorescence correlation or recovery
methods \cite{Pit2000,tretheway2002} have also shown evidence for
the existence of a nonzero slip length.

From a theoretical point of view, the parameters controlling the
magnitude of the slip length are still largely unknown. At a
macroscopic scale, the strength of the interaction between a solid
and a liquid is most obviously characterized through the wetting
behaviour. Weak interactions result in nonwetting behaviour, with
large contact angles for a drop of liquid resting on the solid
substrate. This characterization is  purely thermodynamic, and has
in principle no direct influence on nature of fluid flow past the
interface. Molecular dynamics studies and mode coupling
calculations \cite{Robbins90,Barrat99a,Barrat99b}, however, have
shown that wettability of a perfect surface can be correlated to
the magnitude of the hydrodynamic slippage. Qualitatively, a non
wettable substrate is only weakly coupled to the liquid, with a
depleted surface layer that can be seen as an atomic scale 'air
cushion'. Therefore momentum transfer parallel to the interface is
inefficient, and a large 'slip length' results.

 Experimentally, it appears that although wettability is an important parameter,
different  results can be obtained for substrate/liquid
combinations with similar  wetting properties
\cite{Granick2002,soumis,Pit2000,Stone2002}.  Another parameter of
obvious importance, which may explain such variability, and has
not been taken into account previously in molecular simulations,
is surface roughness. In fact, it was shown by Richardson
\cite{Richardson73} that roughness suppresses slippage on a
macroscopic scale, for any type of microscopic conditions.

Zhu's experiments \cite{Granick2002} have indeed shown that, in a
situation where slippage is observed on smooth surfaces, it can be
suppressed by increasing surface roughness. Other experiments
carried out by Watanabe \textit{et al.} \cite{watanabe99} on
highly water-repellent walls (i.e. walls bearing a  pattern of
narrow parallel grooves) resulted on the other hand on  important
slippage at  the wall. In this work, we present a numerical study
of slippage at interfaces bearing a model roughness, which takes
the form of a nanoscale periodic pattern. The effect of such
patterns on wetting properties has been studied extensively, both
experimentally and theoretically, but their influence on dynamics
has received much less attention.

\begin{figure}[h]
\centering
\includegraphics[height=20cm]{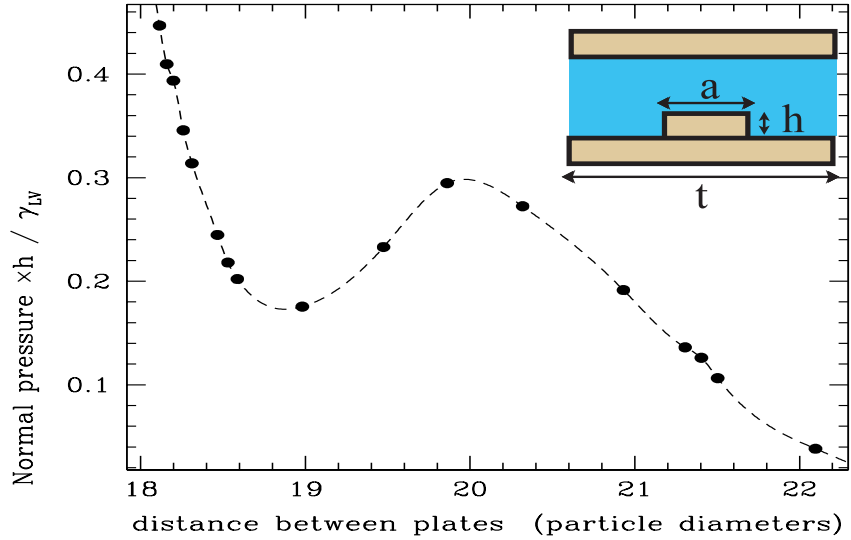}
\vspace{-9cm}
 \label{fig1} \caption{
Normal pressure versus thickness. {}{In order to facilitate
comparison with experimental situations, the pressure has been
rescaled by the typical pressure $\gamma_{LV}/h$ (typically 100
bars for dots of 7nm in water)}. The dashed line is a guide to the
eye. The inset is used to define the main geometrical variables
that describe our setup. The curve shown in the figure corresponds
to $h=5 \sigma$, $t=20 \sigma$, $a=6.6 \sigma$. With these
parameters, a direct application of the macroscopic equation
\ref{pcoex}, with  $\cos\theta=-0.74$ \cite{Barrat99a}, would
yield a coexistence pressure $P_{coex}h/\gamma_{LV}\simeq 0.014$.
Note that the range of distances for which the pressure increases
with $h$ is thermodynamically unstable, its observation being an
artefact of small size simulations.}
\end{figure}

The configuration considered in our  study is a fluid slab
confined between two parallel solid walls. The bottom wall is
decorated with a periodic array of square shaped  dots of height
$h$ and width $a$ (see figure \ref{fig1}). The typical lateral
size of the cell is $L_x=L_y=t=20\sigma$. Periodic boundary
conditions in the directions $x$ and $y$ parallel to the wall are
used. In our simulations,  all the interactions are of the
Lennard-Jones type
\begin{equation} \label{equLJ}
v_{ij}=4\epsilon\left[\left(\frac{\sigma}{r}\right)^{12}-
c_{ij}\left(\frac{\sigma}{r}\right)^6\right]
\end{equation}
The fluid  and the solid atoms have the same molecular diameter
$\sigma$ and interaction energies $\epsilon$. The variation in the
$c_{ij}$ (the index $i,j=F,S$ refers to the fluid or solid phase)
is a convenient  control parameters that can be varied to adjust
the surface tensions. The solid substrate is described by atoms
fixed on a the (100) plane of an FCC lattice. We have worked with
two values of  $c_{FS}$, $c_{FS}=0.5$ or $c_{FS}=0.8$, which
correspond to  contact angles (deduced from  Young's law) of
$\theta=137^{o}$ or $\theta=110^{o}$,
respectively\cite{Barrat99b}. The simulations are carried out at
constant temperature $k_{B}T/\epsilon$=1. In flow experiments,
 the
velocity component in the directions orthogonal to the flow was
thermostatted\cite{Barrat99b}, in order to avoid viscous heating
within the fluid slab. {}{All results reported are obtained within
a linear response regime.} Most numerical results will be given in
Lennard-Jones units (L.J.u.), taking the diameter $\sigma$ as the
unit of length and the energy $\epsilon$ as the unit of energy.

We first briefly describe the static properties of the system,
which were obtained using the following procedure. For a fixed
number of liquid atoms $N_L$ and distance between the planes, the
normal pressure is obtained from the average force on the
substrates along the $z$ direction. This normal pressure can be
varied either by changing the distance between the two walls at a
fixed surface density, or by modifying the number of particles at
fixed distance.

\begin{figure}
\centering
\includegraphics[width=17cm]{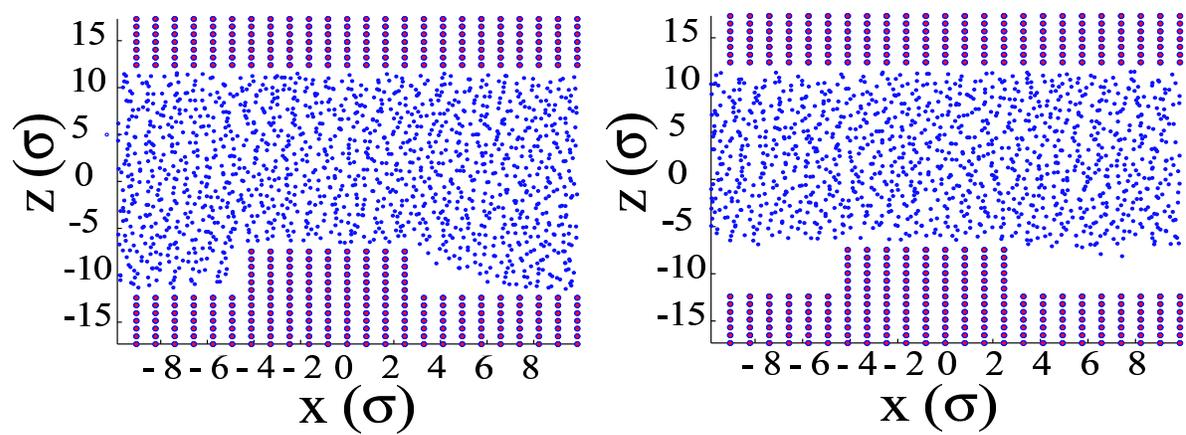}
\caption{Left panel: transverse view of the atomic  configuration
in the 'wetted' situation. Atoms belonging to the liquid and solid
are represented by points and round dots, respectively. The liquid
occupies nearly all the available volume. Right panel: same as
left panel, under conditions where a composite interface is
formed.} \label{confA}
\end{figure}

 Figure \ref{fig1} shows the pressure-thickness
curve obtained when using the first procedure. Two branches
separated by a typical 'van der Waals loop' are clearly visible,
indicating the existence of a phase transition between two
possible situations. At  higher normal pressures, the liquid
occupies all the cell, including the grooves separating the square
dots. At lower pressures, partial dewetting is observed and a
composite interface is formed, the space between the dots being
essentially free from liquid atoms  (see figure \ref{confA}).
{}{Note that the range of pressure in figure \ref{fig1} is rather
small, so that the properties of the bulk fluid over this range
are essentially constant (the relative density change being
typically less than 0.2\%)}

 In spite of the very small sizes
involved, a qualitative interpretation of the observed behaviour
can easily be given in terms of macroscopic capillarity. If we
consider a system at fixed normal pressure $P_ N$, the difference
in Gibbs free energies between wetted configuration (case $A$) and
the formation of a composite interface (case $B$) can be written
as:
\begin{equation}\label{Fdem}
G_B-G_A=(t^2+4ah-a^2).(\gamma_{LV}\cos\theta)+(t^2-a^2).\gamma_{LV}+P_N(t^2-a^2)h
\end{equation}
The composite interface is therefore favoured when
\begin{equation}
P_N<P_{composite}=\frac{-\gamma_{LV}.(\cos\theta+1)}{h}-\frac{4a\gamma_{LV}\cos\theta}{t^2-a^2}
\label{pcoex}
\end{equation}
Although a quantitative agreement can hardly be expected in view
of the small sizes in our system, we have checked that equation
\ref{pcoex} correctly predicts the general trends observed in our
simulation, when the height or width or the square dots are
varied. {}{It can be used, e.g., to understand the influence of an
increase in the corrugation wavelength ($t$ and $a$). At fixed
amplitude $h$, such an increase results in a decrease of
$P_{composite}$, down to unphysically negative pressures. However
dewetting may persist up to large, say micrometric, scales for
adequately chosen asperity sizes \cite{quere2002}.}

 We now consider  the essential objective of our study, namely the
 influence of roughness on dynamical properties of the
confined fluid layers. Our study involved parallel Couette flow,
with the upper wall is moved with velocity $U$ and the lower wall
is moved with a velocity $-U$ (typically $U=0.3$ in reduced
Lennard-Jones units). Thermostatting through velocity rescaling in
the direction perpendicular to the flow is used to keep the
temperature constant.  For flat walls \cite{Barrat99a}, the slip
length $\delta$ is defined as the distance between the wall
position and the depth at which the extrapolated velocity profile
reaches the nominal wall velocity, $v=U$. In the presence of
square dots, the same definition is used. Obviously the presence
of dots makes the choice of the wall position somewhat arbitrary.
We choose to define the wall as the position of the bottom layer
of substrate.

In the following, we will discuss the particular case of a
fluid-solid interaction $c_{FS}=0.5$, which corresponds to a
contact angle $\theta=137^{o}$ on a flat substrate. For this
particular value of the interaction, and in the range of pressures
we have investigated, the fluid displays a moderate amount of slip
at a flat interface, that can be characterized by a slip length
$\delta$ in the range   $20-25\sigma$ (the actual value being
slightly pressure dependent). Taking a molecular size of  $0.5$ to
$1$nm, this corresponds to a value of $10-25$nm.

\begin{figure}
\includegraphics[width=17cm]{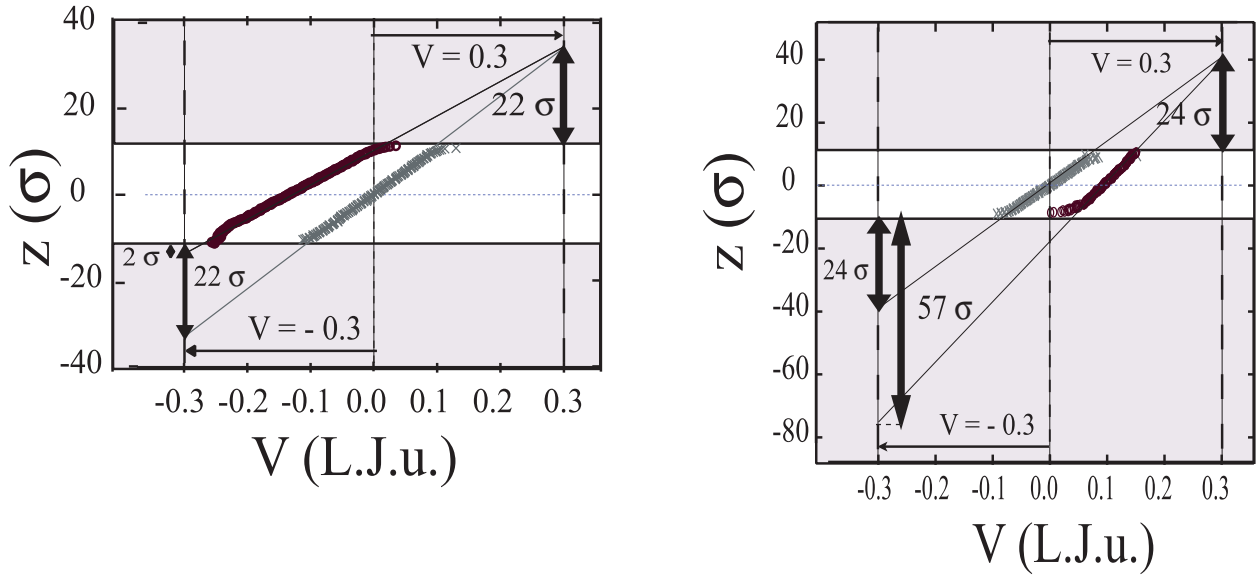}
\caption{Left panel: Flow properties at a sheared solid-liquid
interface, in a wetted situation. The normal pressure is $P=0.086
L.J.u.$.Black dots: velocity profile in the presence of a square
pattern on the bottom wall. The pattern is not represented in the
figure. Grey points: velocity profile with two smooth walls (at
the pressure P=0.086 L.J.u.). The walls are moved at  fixed
velocities $U=\pm0.3$ (in Lennard-Jones units). Slip length are
deduced from the intersections of the velocity profiles with the
axis V=$\pm0.3$. Numerically,  $\delta=22\sigma$ at the flat wall
and $\delta=2\sigma$ at the patterned wall.  Right panel: same as
left panel, for a  composite interface.  The normal pressure is
$P=0.024$ L.J.u.. One now finds, $\delta=24\sigma$ at the flat
wall and $\delta=57\sigma$ at the patterned wall.
 } \label{tuegliss}
\end{figure}

In the presence of the square dot pattern, two very different
situations have to be distinguished. The first case is that  of a
completely wetted substrate (figure \ref{confA}). A typical
velocity profile for this situation is shown in figure
 \ref{tuegliss}. First, it is seen from this profile that the
 introduction of the patterned substrate does not modify  the
 flow in the vicinity of the  upper, structureless wall. This boundary can
 still be characterized by a slip   length $\delta=22\sigma$, identical to what
 would be obtained with two flat substrates at the same pressure.
 In contrast, the slip is strongly suppressed at the lower wall,
 where it can be characterized by a value $\delta\simeq 2\sigma$.
 {}{This value would be slightly higher ($7\sigma$) if the corrugation
 was taken at the crest of the pattern, but would still be much smaller than
 the value obtained for a perfectly flat surface.}

A completely different result is obtained in the case where the
pressure is low enough that a composite interface is formed. A
typical velocity profile corresponding to this situation is shown
in the right panel of figure \ref{tuegliss}. As in the wetted
situation, the pattern on the lower wall does not modify the slip
effect on the upper one. The slip effect on the lower wall is, on
the other hand, strongly enhanced by the presence of the composite
interface. Numerically, the slip length increases by a factor of
about $2.5$, reaching a value  $\delta\simeq 57\sigma$.

Qualitatively, the increase in the slip length should be
associated with the absence of friction at the vapour-liquid
interface, which can be described by a zero stress boundary
condition \cite{Stone2002}. Hence, increasing the liquid vapour
interfacial area by using a more 'spiky' pattern should result in
larger slip length. Using a smaller value of $a$ ($a=4.9 \sigma$)
we find indeed that the slip length can reach a value
$\delta=130\sigma$. {}{It is also interesting to compare these
simulations results with what could be inferred from approximate
hydrodynamic calculations such as that of Hocking
\cite{Hocking75}. Hocking considered in particular flow past a
composite interface, and showed that this results in a partial
slip condition when the fluid filling the corrugation (in our case
vapour)  is of lower viscosity. Unfortunately, a direct comparison
with our results is not possible, due to the use of rather
different corrugation models and to the 'no-slip' condition used
by Hocking at the molecular level. We nevertheless expect that our
results should stimulate further work using continuum approaches}.

Our results confirm that mesoscopic   roughness at the solid
liquid interface can drastically modify the interfacial flow
properties, in the same manner as it affects static wetting
properties. Experimentally, very few results involving surfaces of
controlled roughness are available, and, as mentioned above, these
experiments  yield contrasting  results. We believe this
variability may find its roots in different wetting situations
realized at the mesoscale. It has also been advocated that the
existence of 'nanobubbles' at the liquid solid interface is an
important factor in slippage phenomena. Our simulations, also they
do not, strictly speaking, confirm the existence of such bubbles,
show nevertheless that a composite interface can indeed enhance
slippage considerably. In our simulations, this enhancement
corresponds to an equilibrium situation, which is rapidly achieved
at such small scales. {}{Regarding this point, the situation in
experiments, often done with surfaces bearing a micrometric
pattern, is less clear cut since metastable trapping of bubbles
(possibly made of dissolved gases) may occur into submicrometric
channels}. {}{However, the essential ingredient of the effect
 studied here, i.e. dewetting, is known to be present on patterned
 surfaces, from the nanometer up to the micrometric scale. Therefore, although
 our simulations are performed at nanometer scales, a
 friction reduction is expected over a large range of length scale, up to microscale
patterns. The effect is however expected to be stronger for the more "spiky" nanometer
pattern.}
In fact, very spectacular drops in the
 flow resistance of droplets have been reported \cite{kim2002} on surfaces
 that where decorated with a spiky nanoscale pattern, while the
 effect of a similar microscale pattern was less pronounced.

 In summary, use of {}{patterned surfaces},
and treated to produce a 'water repellent' like effect, appears to be
 a promising way towards devices that would allow flow of liquids
 in small size channels with very small flow resistance.

{\bf Acknowledgments} It is a pleasure to thank Prof. H.A.  Stone
 for interesting discussions. We thank the DGA for its financial
support, and the PSMN (ENS-Lyon) and CDCSP (University of Lyon)
for the use of their computational facilities.

\newpage

\newpage

\newpage

\newpage

\end{document}